\documentstyle[sprocl]{article}
\begin{document}
\vspace*{-2.cm}
\title{CONFINEMENT IN LIGHT-FRONT QCD\footnote{Contributed to PANIC96}}
\author{M.~BURKARDT}
\address{Department of Physics,New Mexico State University \\
Las Cruces, New Mexico 88003-0001 USA}
\maketitle
\begin{abstract}
Numerical results for the calculation of the (rest-frame)
$Q\bar{Q}$ potential, for light-front quantized
$QCD_{2+1}$ on a $\perp$ lattice are presented.
Both in the longitudinal as well as the $\perp$ spatial
directions one obtains linear confinement. The resulting
potential is almost rotational symmetric.
\end{abstract}
Light-front (LF) quantization is the most physical approach to
calculating parton distributions on the basis of QCD~\cite{adv}.
In the transverse lattice formulation of QCD~\cite{bardeen}, 
one keeps the
time direction and one spatial direction continuous and discretizes
the transverse spatial directions. The space time geometry is thus
an array of $1+1$ dimensional sheets.
One the one hand, this provide
both UV and IV cutoffs in the transverse directions and on the
other hand one can still perform LF quantization since the
longitudinal directions are continuous. Furthermore, in the
compact formulation, it is straightforward to implement Gauss'
law as a constraint on the states, which helps avoid troublesome
divergences for $k^+\rightarrow 0$, which plague many other formulations
of LF QCD.

Another advantage of the transverse lattice is that confinement
is manifest in the limit of large $\perp$ lattice spacing $a_\perp$.
The mechanism differs for the longitudinal and the $\perp$
directions: if one separates a $Q\bar{Q}$ pair longitudinally
then, since $a_\perp$ is large, the fields in different sheets
couple only weakly and the quarks interact only with fields
in the same sheet, i.e. effectively the theory reduces to 
$1+1$ dimensional QCD, where confinement is known to be linear.
In contrast, when one separates the $Q\bar{Q}$ pair transversely,
gauge invariance demands that they are connected by a chain of
(gluon) link fields. For large $a_\perp$, where there are only
little fluctuations, this implies that the energy of such a 
configuration is given by the energy for creating one link
quantum times the number of link quanta, i.e. linear confinement
also in the $\perp$ direction. Since the confinement mechanisms
are very different for these two cases, one might ask whether
a rotationally invariant $Q\bar{Q}$ potential results in the
continuum limit. In fact, in the limit of large $a_\perp$ one
finds in $2+1$ dimensions\cite{conf:prd}:
$V(x_L,x_\perp) = \sigma \left(|x_L|+|x_\perp|\right),$
where $\sigma$ is the string tension, which is clearly not
rotationally invariant.
In order to investigate this issue, I used DLCQ and a Lanczos
algorithm to calculate
the rest frame $Q\bar{Q}$ potential from the LF Hamiltonian
for $QCD_{2+1}$ on a $\perp$ lattice. The procedure for computing
the rest frame potential in this formalism follows Ref.\cite{zako}.
An approximation, where one allows at most one link field
quantum per link, was used. Within this approximation, one obtains
only a first order phase transition at the critical point, i.e.
the lattice spacing always remains finite in physical units.
The calculations were done at the critical point, where
$a_\perp \approx 1.04 \sigma^{-1/2}$. 
The resulting $Q\bar{Q}$ potential is shown in the Figure as a 
function of $r =\sqrt{x_\perp^2+x_L^2}$.
\unitlength1.cm
\begin{picture}(15,9)(0,1)
\includegraphics{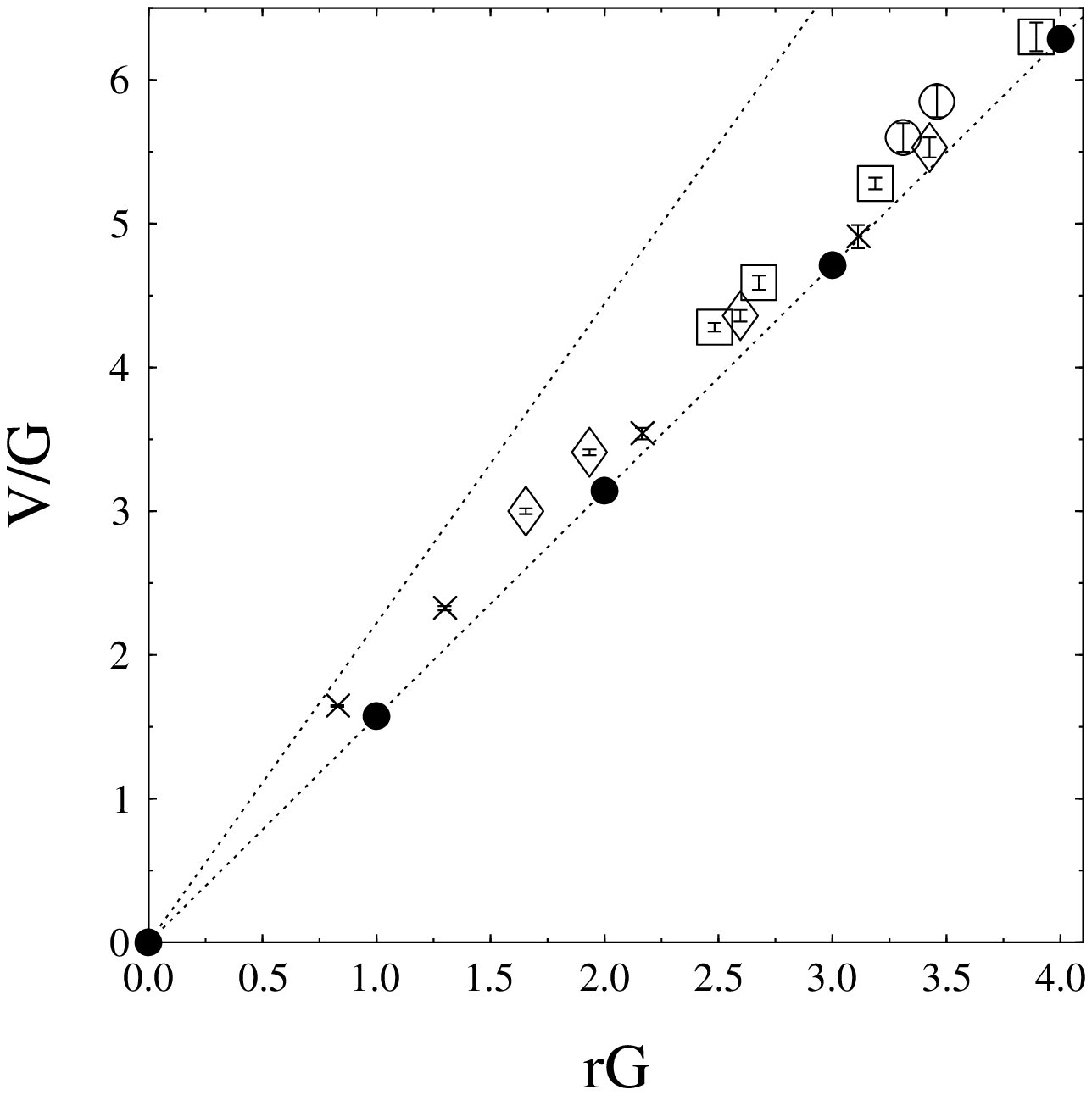}
\end{picture}
One free dimensionless parameter was adjusted in the calculation 
to get equal string tensions in the longitudinal
and transverse directions. For large $a_\perp$, $V(x_\perp,x_L)$
fills the area between the dotted lines. 
The symbols are numerical results
at the critical point for various directions.
The small residual anisotropy is probably due to the truncation
to one quantum per link and due to the simplified ansatz (quadratic)
for the
effective link-field potential.


\section*{References}
\begin{thebibliography}{99}
\bibitem{adv} M.~Burkardt, Advances Nucl. Phys. {\bf 23}, 1 (1996).
\bibitem{bardeen} W.~A.Bardeen et al., 
Phys. Rev. D {\bf 21}, 1037 (1980).
\bibitem{conf:prd} M.~Burkardt and B.~Klindworth, hep-ph/9601289.
\bibitem{conf:elfe} M.~Burkardt, ELFE Meeting 1995, hep-ph/9510264.
\bibitem{zako} M.~Burkardt, LF-workshop,
Zakopane, Aug 1994, hep-ph/9410219.
\end{thebibliography}
\end{document}